\documentclass[11pt,epsf,letterpaper]{article}%
\usepackage{color}
\usepackage{amsmath}
\usepackage{amsfonts}
\usepackage{verbatim}
\usepackage{amssymb}
\usepackage{graphicx}%
\providecommand{\U}[1]{\protect\rule{.1in}{.1in}}
\begin{document}

\title{\flushright{\small UAI-2012-024} \\\center{Black Holes with Primary Hair in gauged N=8 Supergravity.}}
\author{Andr\'{e}s Anabal\'{o}n$^{1}$, Fabrizio Canfora$^{2,3}$, Alex Giacomini$^{4}$
and Julio Oliva$^{4}$\\$^{1}$\textit{Departamento de Ciencias, Facultad de Artes Liberales, Facultad
de}\\\textit{Ingenier\'{\i}a y Ciencias, Universidad Adolfo Ib\'{a}\~{n}ez,
Vi\~{n}a Del Mar, Chile.}\\$^{2}$\textit{Centro de Estudios Cient\'{\i}ficos (CECS), Casilla 1469,
Valdivia, Chile.}\\$^{3}$\textit{Universidad Andres Bello, Av. Republica 440, Santiago, Chile.}\\$^{4}$\textit{Instituto de Ciencias F\'isicas y Matem\'aticas,}\\\textit{Facultad de Ciencias, Universidad Austral de Chile, Valdivia, Chile.}\\{\small andres.anabalon@uai.cl, canfora@cecs.cl, alexgiacomini@uach.cl,
julio.oliva@docentes.uach.cl}}
\maketitle

\begin{abstract}
In this paper, we analyze the static solutions for the $U(1)^{4}$ consistent
truncation of the maximally supersymmetric gauged supergravity in four
dimensions. Using a new parametrization of the known solutions it is shown
that for fixed charges there exist three possible black hole configurations
according to the pattern of symmetry breaking of the (scalars sector of the)
Lagrangian. Namely a black hole without scalar fields, a black hole with a
primary hair and a black hole with a secondary hair respectively. This is the
first, exact, example of a black hole with a primary scalar hair, where both
the black hole and the scalar fields are regular on and outside the horizon.
The configurations with secondary and primary hair can be interpreted as a
spontaneous symmetry breaking of discrete permutation and reflection
symmetries of the action. It is shown that there exist a triple point in the
thermodynamic phase space where the three solution coexist. The corresponding
phase transitions are discussed and the free energies are written explicitly
as function of the thermodynamic coordinates in the uncharged case. In the
charged case the free energies of the primary hair and the hairless black hole
are also given as functions of the thermodynamic coordinates.

\end{abstract}

%\pacs{04.20.Jb 04.40.Nr 04.70.Bw }

%<<<<<<<<<<<<< TITLE >>>>>>>>>>>>>>>%

%<<<<<<<<<<<<< AUTHOR >>>>>>>>>>>>>>>%

%<<<<<<<<<<<<< ADDRESS >>>>>>>>>>>>>>>%

%<<<<<<<<<<<<< DATE >>>>>>>>>>>>>>>%

%======================================%
%<<<<<<<<<<<<< ABSTRACT >>>>>>>>>>>>>>>%
%======================================%

%<<<<<<<<<<<<< PACS NUMBER >>>>>>>>>>>>>>>%

%======================================%
%<<<<<<<<<<<< SECTION I  >>>>>>>>>>>>>>%
%======================================%

\section{Introduction and summary}

Just a few years after coining the term \textquotedblleft Black
Hole\textquotedblright\ J. A. Wheeler, motivated by the results on the
uniqueness of static \cite{Israel:1967wq} and stationary black holes
\cite{Carter:1971zc}, proposed the no hair conjecture \cite{rw1971}. While the
black hole uniqueness theorems states that given the mass, the electric charge
and the angular momentum of an asymptotically flat, four dimensional
configuration, which is regular outside and on the event horizon, it is
possible to single out the Kerr metric in the vacuum case or the Kerr-Newman
in the electrovac case (for references and a nice historical account see the
review by Robinson \cite{Robinson:2004zz}. A comprehensive introduction to the
subject can be found in the book by Heusler \cite{Heusler:1996}), the no-hair
theorem states that it is not possible to endow neither to deform these
space-times, with regular configurations of matter and therefore they would be
the unique final state of gravitational collapse observed in nature;
certainly, a remarkable conclusion. The conjecture was later extended to the
form of theorems in a number of cases, being disproved in the
Einstein-Yang-Mills case and Einstein-Skyrme theory (for a modern discussion
and references see \cite{Hertog:2006rr}) and revitalized by the
Bocharova-Bronnikov-Melnikov-Bekenstein black hole for a conformally coupled
scalar field
\cite{Bocharova-Bronnikov-Melnikov,Bekenstein:1975ts,Virbhadra:1993st}. The
four dimensional asymptotically flat case keep attracting the attention of the
community on both their theoretical \cite{AyonBeato:2002cm} and observational
aspects, see for instance \cite{Johannsen:2010xs}.

The picture was drastically changed with the inclusion of either higher
dimensions or a cosmological constant. Indeed, the very existence of the black
ring \cite{Emparan:2001wn} shows that black holes are not completely
characterized by boundary conditions plus conserved charges in dimensions
higher than four, and the situation is even more striking in the case of the
black Saturn \cite{Elvang:2007rd} since the condition of vanishing net angular
momentum makes the five-dimensional Schwarzschild black hole non-unique.
Actually, in five dimensions, it is known to exist an infinite non-uniqueness
of black rings coupled to a two form potential. Since the total dipole charge
vanishes, the solution it is labeled by a continuous parameter not related
with any conserved charge and it can be regarded as a class of primary hair
\cite{Emparan:2004wy}.

On the other hand, the Maldacena conjecture turned the attention to the case
when the space-time is asymptotically of negative constant curvature and a
number of new hairy black holes were found \cite{Martinez:2002ru}. The
existence of this hairy black holes, in four dimensions, allowed the discovery
of a phase transition when the boundary is in the conformal equivalence class
of $%
%TCIMACRO{\U{211d} }%
%BeginExpansion
\mathbb{R}
%EndExpansion
\times H^{2}$ \cite{Martinez:2004nb} in one of the Duff-Liu black holes (which
are nicely described in section 6.6 of \cite{Duff:1999rk}). When the conformal
equivalence class of the boundary is $%
%TCIMACRO{\U{211d} }%
%BeginExpansion
\mathbb{R}
%EndExpansion
^{3}$ and a $U(1)$ vector is included in the system, the second order phase
transition can be interpreted as the holographic dual of superconductivity
\cite{Hartnoll:2008vx, Koutsoumbas:2009pa}.

However, all the black holes that have been constructed until know, in four
dimensions, do not defy the spirit of the original no-hair theorems. The value
of the scalar hair is actually proportional to the conserved charges of the
system and therefore the scalar field does not have its own integration
constant. This means that there is no continuous degeneration in the
configuration, namely the scalar hair it is switched on or it is not. This
kind of hair have been baptized under the name of \textit{secondary hair}. To
the best of our knowledge, in four dimensional supergravity there are no known
exact\footnote{There is numerical evidence of their existence in the case
where the topology of the horizon is a sphere \cite{Hertog:2004dr}. We found
that the spherically symmetric, uncharged, hairy black holes are singular.}
regular configurations with an independent parameter associated to the scalar
field, such that its value can be arbitrarily varied keeping the conserved
charges fixed, namely \textit{primary hair}. The main objective of this paper
it is to show that such configurations exist. In particular, in the $U(1)^{4}$
consistent truncation of the gauged $\mathcal{N}=8$ supergravity (for a review
and references to the long and interesting history of the gauged
$\mathcal{N}=8$ supergravity theory see \cite{Duff:1999rk}) one can construct
a solution with primary hair.

The key to the construction of a solution with primary hair is the analysis of
the discrete symmetries of the Lagrangian and the possible pattern of breaking
of such symmetries. The Lagrangian of the $U(1)^{4}$ consistent truncation of
the gauged $\mathcal{N}=8$ supergravity possesses two discrete symmetries in
the scalar fields sector: namely a permutation symmetry of the scalar fields
and a reflection symmetry. The hairless case can be considered as a classical
solution in which both symmetries are unbroken while the secondary hair
corresponds to a spontaneous breaking of the reflection symmetry whereas the
permutation symmetry remains unbroken. Thus, one may wonder whether there is a
solution which breaks the permutation symmetry as well. Indeed, as it will be
shown in what follows, such a solution is precisely the sought primary hair.

The paper is organized as follows. In the first section the Duff-Liu static
solutions of the gauged $\mathcal{N}=8$ supergravity are given with a new
parametrization for the conserved quantities. The Ashtekar-Das-Magnon (AMD)
mass \cite{Ashtekar:1984zz} of these configurations in this parametrization is
computed. With the mass at hand it is obvious from the parametrization
previously introduced that there is a degeneration in the phase space of the
theory. Namely, there is more than one black hole for each value of the
conserved charges. In particular, it is possible to see that there are three
possible uncharged configurations. While two of the uncharged ones are well
known, the third is new and has the peculiarity of having vanishing mass and
non-vanishing scalar field. By comparing the curvature scalars of this hairy
black holes we conclude that there exists a continuous parameter characterizing
these configurations, which can be varied without affecting the value of the
conserved charges; thus, defining a primary hair.

In the second section, the thermodynamic properties of the three uncharged
black holes are discussed. The free energy of the three configurations at
fixed temperature is compared and it is shown that there exist a critical
temperature at which all the configurations coexist. In the third section the
previous results are extended to the charged case.

The notation follows \cite{wald}. The conventions of curvature tensors are
$[\nabla_{\rho},\nabla_{\sigma}]V^{\mu}={R^{\mu}}_{\nu\rho\sigma}V^{\nu}$ and
$R_{\mu\nu}={R^{\rho}}_{\mu\rho\nu}$. The metric signature is taken to be
$(-,+,+,+)$, Greek letters are space-time indices and we set $c=1=16\pi G$.

%======================================%
%<<<<<<<<<<<< SECTION I  >>>>>>>>>>>>>>%
%======================================%

\section{The configuration and its charges}

The theory considered here is the consistent truncation of $\mathcal{N}=8$
gauged supergravity \cite{DUFF} described by the action:%

\begin{equation}
S=\int d^{4}x\sqrt{-g}\left(  R+\sum\limits_{m=1}^{3}\left[  -\frac{1}%
{2}\left(  \partial\varphi_{m}\right)  ^{2}+\frac{2}{l^{2}}\cosh(\varphi
_{m})\right]  -\frac{1}{4}\sum\limits_{i=1}^{4}e^{\vec{a}_{i}\cdot\vec
{\varphi}}\left(  F^{(i)}\right)  ^{2}\right)  , \label{Action}%
\end{equation}%
\begin{equation}
\vec{\varphi}=(\varphi_{1},\varphi_{2},\varphi_{3}),\qquad\vec{a}%
_{1}=(1,1,1),\qquad\vec{a}_{2}=(1,-1,-1),\qquad\vec{a}_{3}=(-1,1,-1),\qquad
\vec{a}_{4}=(-1,-1,1). \label{aes}%
\end{equation}
In the sector in which the gauge fields vanish ($F^{(i)}=0$ for $i=1,..,4$),
this action possesses two discrete symmetries $\varphi_{m}\rightarrow
\varphi_{\pi(m)}$ (where $\pi(m)$ is an arbitrary permutation of $(1,2,3)$)
and $\varphi_{m}\rightarrow-\varphi_{m}$ (reflection). The general static
black hole solution can be written by introducing new parameters $Q_{i}$
($i=1,..,4$), such that:
\begin{align}
ds^{2}  &  =-\frac{f}{\sqrt{H}}dt^{2}+\sqrt{H}\left(  \frac{dr^{2}}{f}%
+r^{2}d\sigma_{k}^{2}\right)  ,\qquad f=k+\frac{r^{2}}{l^{2}}H-\frac{\mu}%
{r},\qquad H=H_{1}H_{2}H_{3}H_{4},\qquad H_{i}=1+\frac{Q_{i}\mu}%
{r},\label{BH1}\\
\varphi_{1}  &  =\frac{1}{2}\ln(\frac{H_{1}H_{2}}{H_{3}H_{4}}),\qquad
\varphi_{2}=\frac{1}{2}\ln(\frac{H_{1}H_{3}}{H_{2}H_{4}}),\qquad\varphi
_{3}=\frac{1}{2}\ln(\frac{H_{1}H_{4}}{H_{2}H_{3}}),\qquad A^{(i)}=\pm
\frac{\sqrt{Q_{i}+kQ_{i}^{2}}\mu}{r+Q_{i}\mu}dt. \label{BH4}%
\end{align}

Here $d\sigma_{k}^{2}$ is the line element of a compact manifold of constant
curvature $k$ normalized to $k=\pm1,0$. This new parametrization is related to
the one of Eq. (6.40) and (6.41) of \cite{Duff:1999rk} through $Q_{i}%
=\sinh(\beta_{i})^{2}/k$. The AMD mass \cite{Ashtekar:1984zz} has been used to
compute the conserved charges of a large family of rotating black holes in
gauged supergravity in \cite{Chen:2005zj}. In the present case the AMD mass
and the electric charges of the black holes (\ref{BH1})-(\ref{BH4}) are:
\begin{align}
M  &  =\frac{\mu(2+k(Q_{1}+Q_{2}+Q_{3}+Q_{4}))}{16\pi}\sigma\ ,\label{mass}\\
q_{i}  &  =\frac{1}{16\pi}\lim_{r\rightarrow\infty}\int_{\sigma}e^{\vec{a}%
_{i}\cdot\vec{\varphi}}\ast F^{(i)}=\frac{\sqrt{Q_{i}+kQ_{i}^{2}}\mu\sigma
}{16\pi}\ , \label{charge}%
\end{align}
where $\sigma$ is the volume of the base manifold\footnote{The relation
between the volume of the compact negative constant curvature manifold and its
genus is given by $\sigma= 8\pi(g-1)$ where $g\geq2$}. Note that when $k=-1$,
for each value of the charges $(M,q_{i})$ there is more than one regular black
hole defined by the constants $\mu$ and $Q_{i}$. In other words, for a given
value of the charges there are different black hole configurations and this
leads, as it will be discussed in the next sections, to the appearance of
interesting phase transitions.

Now it is easy  to disclose the \textit{non-uniqueness} mentioned above by
analyzing the uncharged case ($q^{(i)}=0,$ $\forall i$). It is easy to see form
Eqs. (\ref{mass}) and (\ref{charge}) that if one requires that all the charges
are vanishing various possibilities arise:%
\begin{equation}
q_{i}=0\ \ \forall i\Rightarrow Q_{i}-Q_{i}^{2}=0\Rightarrow Q_{i}%
=0\ \ \ or\ \ \ Q_{i}=1\ .
\end{equation}

Thus, it is possible to see that the Duff-Liu family of black holes contains
three \textbf{different} uncharged black holes:%

\begin{align}
\mathbf{I)}\ \ \vec{Q}  &  =\vec{0}\Longrightarrow h=-1+\frac{r^{2}}{l^{2}}%
-\frac{\mu}{r},\qquad\varphi_{i}=0\ ;\label{A}\\
\mathbf{II)}\ \ \vec{Q}  &  =(1,0,0,0)\Longrightarrow h=\left(  -1+\frac
{r^{2}}{l^{2}}\right)  \sqrt{1+\frac{\mu}{r}},\qquad\varphi_{i}=\frac{1}{2}%
\ln(1+\frac{\mu}{r})\ ;\label{B}\\
\mathbf{III)}\ \ \vec{Q}  &  =(1,1,0,0)\Longrightarrow h=-1+\frac{r^{2}}%
{l^{2}}(1+\frac{\mu}{r}),\qquad\varphi_{1}=\ln(1+\frac{\mu}{r})\text{, }%
\qquad\varphi_{2}=\varphi_{3}=0\ ; \label{C}%
\end{align}
where%
\begin{equation}
\vec{Q}=(Q_{1},Q_{2},Q_{3},Q_{4})\ ,\ \ \ h=\frac{f}{\sqrt{H_{1}H_{2}%
H_{3}H_{4}}}\ .
\end{equation}
Any other choice of $\vec{Q}$, corresponds to a relabeling of the scalars or
to the change $\varphi\rightarrow-\varphi$. As it has been anticipated in the
introduction, the three solutions in Eqs. (\ref{A}), (\ref{B}) and (\ref{C})
correspond precisely to the three possible breakings of the discrete
symmetries of the scalars sector of the Lagrangian in Eq. (\ref{Action}).

The black hole in Eq. (\ref{A}) is the usual topological Schwarschild-AdS
space-time with a locally hyperbolic horizon \cite{Mann}: in this case the solutions does
not break neither the permutation nor the reflection symmetries of the scalars
sector of the action. On the other hand, the solution in Eq. (\ref{B}) is the
MTZ\footnote{It is easy to see that the change of coordinates $r=\frac
{\rho^{2}}{\rho+M}$ and redefinition $\mu=4M$, brings the configuration
(\ref{B}) to the same form in which is written in \cite{Martinez:2002ru}.
Thus, the MTZ black hole can seen as classical solution of the consisitent
truncation of $\mathcal{N}=8$ supergravity.} black hole \cite{Martinez:2002ru}
in which a third order phase transition occurs \cite{Koutsoumbas:2009pa}. This
solution does not break the permutation symmetry of the scalar
sector\footnote{In this case the three component of the scalar multiplet are
equal and so $\varphi_{m}=\varphi_{\pi(m)}$ for any permutation $\pi(m)$.} but
it breaks the reflection symmetry since the components of the scalar multiplet
are non-vanishing and so $\varphi_{i}\neq-\varphi_{i}$. The integration
constant $\mu$ which gives the \textquotedblleft strength\textquotedblright%
\ of the scalar field, is proportional to the conserved charges, and therefore it
can be interpreted as a secondary hair.

Finally, the black hole (\ref{C}) has vanishing mass. Therefore the value of
the scalar hair, through the parameter $\mu$, can be freely adjusted without
modifying the conserved charges of the spacetime and it thus becomes a primary
hair. Furthermore, from the value of the scalar curvature:
\begin{equation}
R=-\frac{(3r\mu^{3}+27r^{2}\mu^{2}+\mu^{2}l^{2}+48r^{3}\mu+24r^{4})}%
{2(r+\mu)^{2}r^{2}l^{2}}\ ,
\end{equation}
it is possible to see that the change of coordinate that eliminates the
parameter $\mu$ from the scalar field, $r=\mu\rho$, does not have the same
effect in the scalar curvature. We therefore conclude that it is a genuine
dynamical parameter; namely a genuine primary hair. The mechanism presented here has been
firstly suggested in \cite{Anabalon:2011bw}, were it has been shown that, in
the Lovelock theories of gravity, static configurations with vanishing mass
support pure gravitational hairs.

As far as the thermodynamical behavior is concerned, standard arguments from
statistical mechanics suggest that the configurations with broken symmetries
can prevail at low temperatures while one should expect that, at high enough
temperatures, the symmetries should be restored. In the following sections we
will show that this is indeed the case.

\section{Thermodynamics for the uncharged case}

The characterization of the thermodynamics of the configurations discussed in
the previous section is very clear in the canonical ensemble. Since the
gravitational part of the Lagrangians is the Einstein-Hilbert term, it follows
that the black holes satisfy the area law. Requiring the Euclidean
continuation of the black holes, obtained after performing the Wick rotation
$\tau=it$, to be everywhere regular fixes the temperature as the inverse of
the period of the Euclidean time $\tau\in\lbrack0,\beta)$. In this approach
the free energy is defined through the relation $F=M-TS-\Sigma_{i}\Phi
^{(i)}q^{(i)}$, where $\Phi^{(i)}$ are the electric potentials for each of the
Maxwell fields $A^{\left(  i\right)  }$. Since we are considering the
uncharged case ($q^{(i)}=0$, $\forall i$), the free energy reduces to
$F=M-TS$, which is related to the Euclidean action $I$ by $I=-\beta F$. In
what follows the quantities associated to the black hole without scalar field,
with primary hair and with secondary hair are denoted with a subscript
\textquotedblleft$0$\textquotedblright, \textquotedblleft$1$\textquotedblright%
\ and \textquotedblleft$2$\textquotedblright\ respectively. For each of these
configurations, the relevant thermodynamical quantities are:%

\begin{equation}
T_{0}=\frac{2r_{+}+3\mu}{4\pi r_{+}^{2}},\qquad S_{0}=\frac{\sigma r_{+}^{2}%
}{4},\qquad F_{0}=-\frac{\sigma\left(  2r_{+}+\mu\right)  }{16\pi}, \label{A2}%
\end{equation}

\begin{equation}
T_{1}=\frac{\sqrt{\mu^{2}+4l^{2}}}{4\pi l^{2}},\qquad S_{1}=\frac{\sigma
l^{2}}{4}\qquad F_{1}=-\frac{\sigma Tl^{2}}{4}, \label{B2}%
\end{equation}

\begin{equation}
T_{2}=\frac{\sqrt{l+\mu}}{2\pi l^{\frac{3}{2}}},\qquad S_{2}=\frac{l^{2}%
\sqrt{1+\frac{\mu}{l}}\sigma}{4},\qquad F_{2}=-\frac{(2l+\mu)\sigma}{16\pi}.
\label{C2}%
\end{equation}
It is worth noting that the parameter $\mu$ has a different meaning in each of
these black holes. In order to decide which are the configurations favoured
thermodynamically one should compare the corresponding free energies. However,
the above form in which the free energies are seen as functions of $\mu$ and
$r_{+}$ (which are not suitable thermodynamical coordinates) is not
convenient. The physical reason is that when one has to decide which one of
two or more configurations prevails from the thermodynamical point of view one
has to compare the corresponding free energies at the same temperature and
voltage. Therefore, the most convenient procedure is to express the masses and
the entropies in terms of the corresponding temperatures in the three cases
above. This can be done explicitly in the uncharged case. Thus, the free
energies in terms of the temperature read%
\begin{equation}
F_{0}=-\frac{(2T^{2}\pi^{2}l^{2}+T\pi l\sqrt{4T^{2}\pi^{2}l^{2}+3}+3)\sigma
l(2T\pi l+\sqrt{4T^{2}\pi^{2}l^{2}+3})}{108\pi}\ , \label{A3}%
\end{equation}%
\begin{equation}
F_{1}=-T\frac{\sigma l^{2}}{4}\ , \label{B3}%
\end{equation}%
\begin{equation}
F_{2}=-\frac{(l+4T^{2}\pi^{2}l^{3})\sigma}{16\pi}\ . \label{C3}%
\end{equation}
According to our convention, the configuration with less free energy is
thermodynamically the most favorable. It can also be shown that at the
critical temperature
\begin{equation}
T_{c}=\frac{1}{2\pi l}\ ,
\end{equation}
the three free energies coincide and so, around $T=1/(2\pi l)$, the three
phases coexist since they are equally probable. Below the critical point the
secondary hair (which breaks reflections symmetry but does not break the
permutation symmetry) is favored. On the other hand, for temperatures larger
than $T_{c}$ the hairless solution is favored and the symmetry is restored.
The transition from the hairless solution to the secondary hair is the one
reported in \cite{Martinez:2002ru} in the special case of zero electric
charge. As it has been already mentioned, it was shown in
\cite{Koutsoumbas:2009pa} that this transition associated to the breaking of
the reflection symmetry must be of third order.

The terminology used here for the classification of phase transitions is the
one of Ehrenfest where \textquotedblleft\textit{n-th order phase
transition}\textquotedblright\ means that the n-th derivative of the free
energy is discontinuous at the critical point. Many interesting examples of
transitions of order higher than two have been found such as the
Gross-Witten-Wadia third order phase transition \cite{grosswitten}
\cite{thirdorder2} in a unitary matrix model related to lattice QCD (a nice
analysis of the properties of higher order phase transitions is \cite{review}).

In the present case, it can be seen directly\footnote{The advantage
of having an expression for the free energies which only depend on the
thermodynamical coordinates such as temperature and chemical
potentials (as in Eqs. (\ref{A3}), (\ref{B3}) and (\ref{C3})) is
that in order to determine the order of the phase transition one
does not need to perform any Taylor expansion. The phase transition
can be determined in an exact manner.} from Eqs. (\ref{A3}), (\ref{B3})
and (\ref{C3}), the free energies are continuous at the critical
temperature up to their second derivative, and the third derivative
has a discontinuity at $T_{c}$. The primary hair, which is associated
to a spontaneous breaking of both discrete symmetries of (the
scalars sector of) the action, is never favored except at the
critical point where it can coexist with the other two phases. This
is a quite interesting results since it has been previously argued that primary hairs should be unstable \cite{Mavromatos}, thus, one could
have expected a finite gap among the free energy of the primary hair and
the free energies of the other configurations at all the
temperatures. However, we have shown that at $T=T_{c}$ the three
free energies coincide up \textit{to the second derivatives}: this
implies that, close to $T_{c}$, also the primary hair can be
relevant as far as the thermodynamical behavior is concerned.

\section{The charged case}

It is interesting to study whether the picture described above survives when
the black holes support electric charge. The simplest case is when all the
charges are equal: $q_{i}=q$. This implies that either $Q_{i}=Q_{+}$ or
$Q_{i}=Q_{-}$ with
\begin{equation}
Q_{\pm}=\frac{1}{2}\pm\frac{1}{2}\sqrt{1-4\frac{p^{2}}{\mu^{2}}}\ ,
\end{equation}%
\begin{equation}
p=\frac{16\pi q}{\sigma}.
\end{equation}
Thus, the following bound on the charge must hold:%
\begin{equation}
p\leq\frac{\mu}{2}\ .
\end{equation}
Note that $Q_{+}>Q_{-}>0$ and therefore the $H_{i}$ in Eq. (\ref{BH1}) are
positive whenever $\mu r>0$. It follows that the curvature scalars are regular
outside of the horizon. In this case there are also three different
configurations (up to changes of signs of the scalar fields) which can be
characterized by the corresponding $Q-$vectors as in the previous section. The
topological Reisner-Nostdr\"{o}m-AdS black hole is characterized by the
following $Q$-vector:
\begin{equation}
\vec{Q}_{0}=(Q_{-},Q_{-},Q_{-},Q_{-})\ .
\end{equation}
On the other hand, the black holes with primary hair and secondary hairs are
characterized by $\vec{Q}_{1}\ $and $\vec{Q}_{2}$ respectively, where%
\begin{align}
\vec{Q}_{1} &  =(Q_{+},Q_{+},Q_{-},Q_{-})\ ,\\
\vec{Q}_{2} &  =(Q_{+},Q_{-},Q_{-},Q_{-})\ .
\end{align}
As in the previous case we have to express the free energies in terms of the
temperature only in order to compare them properly, in the present case one
should express the three free energies in terms of the temperature $T$ and the
voltage $\Phi$ (which are the correct thermodynamical variables in this case).
This can be done explicitly both for the hairless solution and the primary
hair (while the corresponding expression for the secondary hair involves the
roots of a sixth order algebraic equation so that it cannot be written explicitly):%

\[
F_{0}=-\frac{(32T^{2}\pi^{2}l^{2}+4T\pi l\sqrt{64T^{2}\pi^{2}l^{2}%
+48+3\Phi^{2}}+48+3\Phi^{2})\sigma l(8T\pi l+\sqrt{64T^{2}\pi^{2}%
l^{2}+48+3\Phi^{2}})}{6912\pi}\ ,
\]%
\[
F_{1}=-\frac{Tl\sigma(16l^{2}\pi^{3}+4l\pi\sqrt{16l^{2}\pi^{4}+(\frac{\Phi}%
{T})^{2}}+(\frac{\Phi}{T})^{2})}{32\pi\sqrt{16l^{2}\pi^{4}+(\frac{\Phi}%
{T})^{2}}}\ ,
\]%
\[
\Phi_{2}^{\pm}=\frac{p}{x_{+}+\mu Q_{\pm}},\qquad S_{2}=\frac{l\sqrt
{x_{+}(x_{+}+\mu)}}{4\sigma},\qquad F_{2}=-\frac{\left(  \mu+2x_{+}\right)
\sigma}{16\pi}\ ,\qquad\Phi\equiv\sum_{i=1}^{4}\Phi^{(i)}\ ,
\]
where $x_{+}$ is the radial location of the horizon for the secondary hair.
For small enough voltage $\Phi$ one can show that the picture is very similar
to the one of the previous section. Namely, the hairless case is
thermodynamically more favorable at temperatures higher than the critical
temperature while the secondary hair prevails at low temperatures. The
critical temperature is a sort of triple point where the three free energies
coincide. On the other hand, it can be seen that if one increases the value of
the voltage $\Phi$ then, correspondingly, the range of temperatures where the
secondary hair is thermodynamically favored decreases.

\section{Conclusions}

In this paper the general static black hole solution of the consistently
truncated $\mathcal{N}=8$ gauged supergravity has been analyzed. Using a new
parametrization of the solution space, we showed that for fixed mass and
electric charge, in addition to the usual hairless black hole, there exist
both a black hole with a secondary scalar hair and another black hole with a
primary hair. To the best of our knowledge, this is the first known example of
a scalar primary hair in four dimensional (super) gravity. Remarkably, it
turned out that the MTZ secondary hair solution is a special case of these
solutions when the electric charge is equal to zero. The solutions with scalar
hair are associated to a spontaneous symmetry breaking of the discrete
permutation and reflection symmetries (of the scalars sector of) the action.
In the uncharged case, the free energies can be expressed as functions of the
temperature only and it can be shown analytically that the phase transition
between the hairless solutions (which prevails at high temperatures) and the
secondary hair solution (which prevails at low temperatures) is of third
order. The primary hair is thermodynamically disfavored except at the critical
point $T_{c}$ when the three free energies coincide up to the second
derivative. Thus, when $T$ is close to $T_{c}$ the primary hair can coexist
with the other configurations and it could be relevant as far as the
thermodynamics is concerned. The charges can also be included in the analysis
but they do not change the qualitative picture of the uncharged case.

Finally we would like to recall that it was pointed out, in ungauged
supergravity, that the primary hair is not precluded by SUSY and
therefore it would be possible to construct singular solutions by a
duality transformations of the theory \cite{Ortin:1997yn}, and that
the primary hair should be included in the BPS bound. Since the, in
this case regular, uncharged solution with primary hair has also
vanishing mass it can be regarded as an exotic case of an
``extremal'' black hole. It would be therefore interesting to see
whether the proposal to include the primary hair in the BPS bound
can be achieved in this case, and what it would imply.

%======================================%
%<<<<<<<<<< ACKNOWLEDGEMENT >>>>>>>>>>>%
%======================================%

\paragraph{Acknowledgements}

{\small {The authors would like to thank Pedro Alvarez and Henning
Samtleben for interesting discussions. The research of A.A. is
supported in part by the Alexander von Humboldt foundation. This
work is partially supported by FONDECYT grants 1120352, 1110167, and
11090281, and by the CONICYT grant Anillo \ ACT-91 \textquotedblleft
Southern Theoretical Physics Laboratory\textquotedblright. The
Centro de Estudios Cient\'{\i}ficos (CECs) is funded by the Chilean
Government through the Centers of Excellence Base Financing Program
of CONICYT. F. C. is also supported by Proyecto de Insercion CONICYT
79090034.}}

%%%%%%%%%%%%%%%%%%%%%%%%%%%%%%%%%%%%%%%%%%%%%%%%%%%%%

%======================================%
%<<<<<<<<<<<<< REFERENCES >>>>>>>>>>>>>%
%======================================%
%QTO{references}
{}

\end{document}